# The Case for Globally Beneficial Technology

Iason Gabriel and Atoosa Kasirzadeh

To whom do the fruits of advanced technological innovation belong? To their inventors, to the organizations and individuals involved in making such discoveries possible, or to still larger groups of people, potentially encompassing all of humanity? This question sits at the heart of the present investigation. The arguments developed here focus on an expansive reading of the entitlement to benefit from technological breakthroughs: we argue that they should be designed, developed, and distributed in ways that benefit everyone. This central claim, which encompasses technologies such as advanced forms of artificial intelligence, is grounded in an exploration of five moral arguments that involve human rights, beneficence, contingencies of birth, the global tree of knowledge, and global economic justice. Taken together, they underpin the argument for globally beneficial technologies.

> *Concern for the man himself and his fate must always form the chief interest of all technical endeavors... Never forget this in the midst of your diagrams and equations.*
>
> Albert Einstein

## Introduction

The early 1600s saw the invention of a major new life-saving technology. At that time, obstetric forceps were first developed and successfully trialed by the Chamberlens, who were a family of barber-surgeons practicing in London, heralding a potentially dramatic improvement in child and maternal survival rates for complicated births. Before the invention of medical forceps, the best efforts of doctors involved cesarean sections, which were often fatal, efforts to break the mother's pubic bone, or dangerous attempts to terminate late-stage pregnancies. However, for more than three generations, the new instrument was kept a secret and used exclusively by the Chamberlens to bolster their business, with the forceps being concealed in medical bags and under surgical draperies when they were used for operations. It was not until more than a century later, in 1723, that the Flemish surgeon Jan Palfijn publicly demonstrated and published the details of his *mains de fer* ("iron hands"), a similar device that he had independently invented. Within a decade, medical forceps were adopted and became part of standard medical practice, first in Europe and then more widely. Finally, in 1813, a visitor to Woodham Mortimer Hall in Essex noticed a small cork disk, flush against the floor, along with covered screwheads and a trapdoor with small, sunken hinges. Upon further investigation she found three pairs of the original medical forceps hidden beneath the floorboards of the old Chamberlen family home. At this moment it became possible to appraise the full human cost of what had transpired.

In 1959, the development and deployment of life-saving technology took a different path. A Swedish engineer named Nils Bohlin, who worked at the national automotive company, Volvo, invented the modern three-point seat belt. Bohlin had noticed that earlier, simpler versions led to a pattern of serious and sustained injury: upon impact, people would fall forward, with their heads hitting interior surfaces, and the body twisting in ways that caused serious damage to the spine and internal organs. The new seat-belt design halved the overall rate of injury for people unfortunate enough to find themselves in automobile accidents and increased the odds of surviving a more serious crash by one third. After a period of discussion within the company, it was agreed that the technology would be filed with an "open" patent so that it could be freely used by other manufacturers without paying royalties to the inventor. As a result of the obvious improvements the new design brought with it, adoption spread quickly around the world. Reasonable current estimates suggest that the modern seat-belt design has saved somewhere in the region of 250,000 to 350,000 lives in the United States alone (Kahane, 2015; Sauber-Schatz, 2016).

These two inventions show that the way in which the benefits of technological innovation are distributed can be decidedly important for the world and for those who live in it. In the first case, it is possible to identify human choices that led to the continuation of significant preventable death and suffering; in the second case, this outcome was averted, unlocking further human potential. Highly consequential decisions of this kind also arise for transformative general purpose technologies (GPTs), such as the steam engine, electric motor, and semiconductor, which shape the course of human development over decades or even centuries. As documented by Bresnahan and Trajtenberg (1995), GPTs have a number of properties that make them potentially world-changing: they are *pervasive* and perform a function that can often be adapted to different tasks; they are *dynamic* insofar as they make possible temporarily extended sequences of gains; and they spur *complementary innovations* that extend the value of the initial invention and lead to further use. Recent research has suggested that these properties may soon be met by artificial intelligence (AI) (Eloundou, 2024), with computer scientist Andrew Ng stating that, "AI is the new electricity. It will transform every industry and create huge economic value" (Jewell, 2019). The potential for widespread impact leads directly to questions such as how the risk created by such technologies should be managed.[1] It also raises questions about how the benefits that result from advanced technologies should be distributed: To whom do these benefits belong and on what basis?

[1] The term "technology" is used in this paper to denote any reproducible configuration of material resources, scientific or practical knowledge, and operational infrastructure, developed

In this paper, we advance a radical claim: the benefits of advanced technology belong to the world. More precisely, we provide five arguments in support of what we term the Central Claim, which holds that advanced technologies belong to the world in the sense that everyone has a claim to benefit materially from their invention and use.

Certain elements of the Central Claim are worth highlighting. First, it is fully universal insofar as it holds that *everyone* is morally entitled to benefit from advanced technology, not just certain individuals, organizations, nation-states, or some subset of the global population. Indeed, it includes people who may not have made a direct material contribution to the development of these artifacts. Second, the Central Claim is substantial insofar as it holds that everyone is entitled to benefit *materially* from this technology. The Central Claim provides a normative standard against which different kinds of institutional schema or distribution mechanisms ought to be evaluated. The emphasis on "material benefit" means that this standard extends beyond the notion that people only need to be granted the *opportunity* to benefit from advanced technologies or that they should not be actively impeded in doing so. It is also distinct from the claim that they have a right to say how this technology is governed, or that measures ought to be put in place to control the risks—though these considerations are likely to be both true and consequential for many advanced technologies. Third, the reference to "advanced technologies" is intended to capture kinds of breakthroughs that have the potential to be both *transformational* and *beneficial*. A technology is transformational if it has a profound effect on many aspects of life. This property is demonstrated by many historically important GPTs (Mokyr, 2025), and it also holds, in expectation, for a number of future technologies such as advanced AI (Stern, 2025; Manyika, 2026). A technology is beneficial if it can be expected to create enormous positive value, conditional upon the way in which it is distributed and used. If a technology does not have these two properties, and if, by quirk of fate, advanced AI fails to meet this high bar, the Central Claim may not bear directly upon it, but it would still hold for other qualifying technologies.

To support the Central Claim, it is useful to take stock of certain facts about the world. First, material scarcity and extreme poverty continue to be a reality of life for many people worldwide. One in ten people live on the equivalent of less than $3 per day, and over a billion people currently lack access to essential goods and services (World Bank, 2025). Moreover, the effects of extreme poverty are severe: people living in this

---

to perform specific functional tasks or expand human capabilities (Arthur, 2009; MacKenzie & Wajcman, 1999). It encompasses both localized physical artifacts—such as obstetric forceps or three-point seat belts—and macro-level general purpose systems such as electricity and AI (Winner, 2020).

condition often have limited access to education or suffer from ill health and chronic hunger, which are conditions that lead to wasting and stunting, particularly in early life (OPHI, 2025). Second, wealth, income and resources are distributed very unequally at the global level, generating huge inequality in life prospects (Milanovic, 2011; Milanovic, 2015; UBS, 2025). As a direct consequence of this, it matters greatly *where* in the world a person is born. Third, global inequality has a number of causes that include, among other things, the rules that structure the global economic and legal system (Crivelli, 2016; Tørsløv, 2023; Wenar, 2015). Fourth, these inequalities manifest in the world of science and technology (Acemoglu and Johnson, 2023; Jasanoff, 2016), with two billion people currently lacking Internet access (ITU, 2025). As a consequence, new technologies do not enter a world where basic needs are met or where fair outcomes might be said to obtain at a global level. Rather, advanced technologies are born into a world that contains vast unmet material need and often significant levels of background inequality.

To evaluate the strengths and weaknesses of the Central Claim, and to better understand how it intersects with these facts, this investigation proceeds through several stages. Section 1 looks at the relationships between technology, human rights, and essential goods. Section 2 explores the opportunity technology creates for global benefit and the problem of squandered human potential. Section 3 looks at the importance of luck, fairness, and circumstances of birth. Section 4 looks at claims to benefit from technology arising from the "global tree of knowledge," understood as the collective practice of knowledge creation. Section 5 looks at the relevance of global economic institutions and background injustice for the Central Claim. Finally, Section 6 provides a summary of our findings and explores the practical implications of the Central Claim for advanced technologies such as AI.

Ultimately, we find considerable overlap among the arguments, each of which supports the Central Claim in different ways. Some of the arguments considered here are powerful but not fully universal in scope. Others are universal but apply only to certain kinds of technologies or under certain conditions. Thus, rather than a single knock-down argument, what we discover is a set of moral considerations that jointly ground universal claims about who ought to benefit from advanced technologies and why. This wide grounding is, we believe, collectively robust and sufficient to vindicate the Central Claim in the majority of cases. Nonetheless, the scope of this investigation remains bounded in certain ways—due to limitations of space, we are not able to evaluate *all* the arguments that bear upon the truth or falsity of the Central Claim. Nor are we able to provide a detailed roadmap for *how* universal benefit is to be unlocked in practice. This is an empirical question, and one which requires deeper analysis of various institutional considerations, including how to incentivise continued innovation in a way that creates real technological benefits for all. For genuinely transformative technologies, such as advanced AI, it seems likely that a range of approaches are needed—including the open publication of research and release of models, market access that is free at the point of use, state-level service provision and coordinated global action to ensure that these benefits are unlocked and directed appropriately. Yet, before these institutional mechanisms can be evaluated, we must first establish a clear view of the normative destination towards which we must sail. To that inquiry, we now turn.

## 1. The Human Rights and Essential Goods Argument

On December 10, 1948, delegates from 58 nations gathered at the Palais de Chaillot in Paris to sign the Universal Declaration of Human Rights (UDHR). World-weary, still affected by the burdens of war, but determined to rebuke those "barbarous acts which have outraged the conscience of mankind," the delegates affirmed that "all human beings are born free and equal in dignity and rights. They are endowed with reason and conscience and should act towards one another in a spirit of brotherhood" (United Nations General Assembly, 1948). Furthermore, the signatory parties held that everyone is entitled to these rights, "without distinction of any kind, such as race, colour, sex, language, religion, political or other opinion, national or social origin, property, birth or other status." The Declaration affirms that "everyone has the right to life, liberty and the security of person" as well as human rights to freedom of conscience, health, education, and fair treatment before the law.

However, the aspirations of the signatories extended further. Article 27, paragraph two, speaks directly to the global enterprise of science and technology creation. Alongside a commitment to respect the interests of creators, signatories held that "everyone has the right freely to participate in the cultural life of the community, to enjoy the arts and *to share in scientific advancement and its benefits*."[2] The provision was championed by Latin American delegates who viewed access to science as a prerequisite for economic development, supported by Soviet delegations who wanted to include the further specification that science should serve "the cause of peace," and endorsed by the French jurist and co-author of the draft, René Cassin, who viewed science as part of the common heritage of mankind (Morsink, 1999) (see also Section 4). Though the right to science continues to be a somewhat neglected human right, with concerns about political and civil liberties taking center stage, it nevertheless leads to an important question: Do human rights ground a universal claim to benefit from advanced

[2] The later International Covenant on Economic, Social and Cultural Rights (1966) Article 15 further recognizes the right of everyone "to enjoy the benefits of science and its applications" and holds that relevant parties must take steps to ensure the "development and the diffusion of science and culture."

technologies?

To answer this question, we need to say something more about the character of human rights and the way in which they are justified, topics about which there is considerable ongoing philosophical and legal debate. An important entry point for this discussion can be found via the intuition, articulated by the philosopher Thomas Nagel, that "there is some minimal concern we owe to fellow human beings threatened with starvation or severe malnutrition and early death from easily preventable disease" which generates moral responsibilities that we can all recognize "if we are not simply ethical egoists" (2005, 118). This minimal claim about global morality finds fuller expression in the idea of human rights, understood as a set of foundational moral claims or entitlements held by individual human beings against states, corporations, and a variety of other actors.

Although the precise basis of such rights is contested, they are commonly held to be a normative consequence of ascribing dignity, moral autonomy, or value to human life (Griffin, 2008; Nickel, 2007; Pope Leo XIV, 2026; Sen, 2004). Human rights include *negative rights* not to be harmed in certain ways, and *positive rights* to reliably enjoy goods and resources such as adequate nutrition, health, education, and shelter. Relatedly, there are *negative duties* not to harm people in ways that violate their human rights, and there are *positive duties* to assist people who are unable to access the goods and services needed to achieve a minimally decent human life. Although international law now recognizes a potentially expansive list of such claims and duties, including, for example, the right to paid holiday, the Human Rights and Essential Goods argument considered here does not rely upon a "maximalist" understanding of human rights claims. Instead, it holds that there is a more limited conception of human rights, anchored primarily in concern for people's basic needs, capabilities, and core functionings (Nickel, 2007; Shue, 1980), which provides support for the Central Claim.

Human rights are relevant to the design and deployment of advanced technologies, including AI systems, in at least two respects: (1) permissible use and (2) situational necessity. To begin with, human rights serve as an important *constraint* upon the ways in which new technologies may permissibly be used. People have a right to live lives that are free from persecution and political oppression, and in which they enjoy freedom of conscience, religion, assembly, and access to the rule of law, among other things. Technologies should not be used in ways that are at odds with these goals, and applications of AI that pose risks to human rights, such as those enabling more extensive surveillance by authoritarian states, should be restricted.

Human rights also ground valid claims to access, make use of, or benefit from technology in certain situations (Shaver, 2015). This kind of claim arises most often in connection with positive rights, in cases where access to technology is either *instrumentally necessary* or *particularly important* for human rights claims to be met.[3] This condition is met most clearly when a technology, such as free access to a cellular network or to certain kinds of drugs or medicine, is the only way in which to secure human rights, for instance by supporting essential services after a natural disaster or by preventing significant loss of life during an epidemic. In these cases, the urgency of the claim is typically held to be sufficient to generate a duty to provide them, overriding the right of others to maintain exclusive control over the good, service, or technology in question. At the same time, it is not clear that the positive claim to assistance needs to be so tightly bound, either in terms of temporality or to claims about specific causal pathways. Indeed, on many philosophical accounts, the urgency of a moral claim to assistance is a product of its overall weight—or the severity of the underlying situation—rather than its specific time-bound nature (Ashford, 2018; Scanlon, 1975). Moreover, even where technology provision is not the sole means of satisfying human rights claims, it is still often the *best* feasible option, for reasons explored in more detail below.

In the context of advanced technologies such as AI, a salient question is whether they are already needed for human rights to be fulfilled or whether they might become so in the near future. At the present moment, the instrumental necessity condition may not be met by AI for many standard use cases. Indeed, although it is possible to imagine situations in which AI systems are necessary to secure human rights by, for example, helping to anticipate extreme weather events (Nearing, 2024; Camps-Valls, 2025) or solve coordination problems during emergencies (Beduschi, 2022; Haykal, 2025; Lythreatis, 2026), AI technology is not yet important or pervasive enough to be widely subject to this kind of claim. However, this could change if AI takes on more of the properties associated with transformational GPTs and if we endorse a more expansive notion of the relevant connection between human rights and technology.

Indeed, from a systemic vantage point, almost all basic rights, including access to acceptable levels of nutrition, shelter, health, and education, depend upon access to the fruits of technological innovation and production (at least at the societal level). More narrowly, the use of technology to assist those whose human rights go unmet has advantages over the transfer of money—especially overseas development aid with its various conditionalities and distortions (Alesina and Dollar, 2000; Dreher et al.,

[3] Broadly speaking, we can distinguish between necessity and feasibility conditions for an outcome to be acheived. Necessary conditions must be satisfied for any specific outcome to be possible. For example, the presence of a material medium is necessary for sound, as acoustic waves cannot travel through a vacuum. By contrast, feasibility conditions concern what is achievable in practice, given available resources, social realities, and institutional capacities (Southwood, 2018). When we describe technology provision as the “best feasible” way to secure a right, this implies that while theoretical alternatives might exist, they are likely unreachable in the current context.

2024)—or institutional efforts, which often encounter collective action problems or may take decades to enact. Taking the example of telephony in low-income countries as one relevant data point, it is clear that access to new technologies can support subsistence and economic empowerment (Jensen, 2007; World Bank, 2016), improve education and the flow of information (Aker, 2010), support public health goals (Rotondi et al., 2020), and help address emergencies (Zufiria, 2018), all without the challenges or overheads encountered by other forms of intervention. For these reasons, recourse to technology is often a particularly important tool in the armory of those with a responsibility to see that human rights are upheld and respected.[4]

Nonetheless, when viewed as a potential grounding for the Central Claim, the Human Rights and Essential Goods argument encounters three challenges. First, there are concerns about its generality: Does the argument really generate a claim to benefit from advanced technologies *in general* or only from certain kinds of advanced technology (and not others), depending on the type of technology and situation? Second, there are questions about who the relevant duty-bearers are and about how the responsibilities resulting from human rights obligations should be assigned. As our discussion of the UDHR suggests, the obligation to uphold these rights was initially thought to fall primarily, if not exclusively, upon states. Against this backdrop, one might wonder whether it is really *technologists* who are responsible for discharging these duties even when there are valid human right claims at stake.[5] Third, the level of support it provides for the Central Claim appears to be "capped" in certain ways. That is, while the Human Rights and Essential Goods argument provides strong support for the transfer of technology or technological wealth in situations where there is significant unmet material need, this claim seems to lose its force *beyond a certain point*—and to run out altogether once underlying needs are met.

Each objection has some weight. First, the question of whether there is a fully general claim to benefit from transformational GPTs turns in large part upon how tight the connection needs to be between a human right and conditions for its fulfillment, and whether we adhere to the stronger necessity condition or to the more expansive notion of "importance" developed herein. As the signatories to the UDHR were keenly aware, access to certain kinds of technology is necessary for rights fulfillment in the modern world. Without access to these technologies, efforts to secure the object of these rights on a universal basis would foreseeably and predictably come up short. Developing this argument further, it has often been noted that what a person may justifiably lay claim to via their basic rights is, in part, a product of the era in which they live and the kind of minimally decent life that it makes sense to hope for given the technologies of that age (Nickel, 2007; Tasioulas, 2015). If advanced technologies such as AI make it easier to achieve a higher level of health and education moving forward, the content of these human rights may expand to include more as a matter of basic entitlement (Prabhakaran et al., 2022), and the scope of the Central Claim would similarly expand. Conversely, it may be possible to imagine advanced technologies that do not have such widespread or general impact. The relevance of more narrow scientific discoveries for human rights fulfillment would then be more constrained, dampening the force of the Central Claim for this subset of cases. Thus, the Human Rights and Essential Goods argument provides greatest support for the Central Claim in relation to the transformational GPTs that are also its main object.

---

[4] This is quite different from the "techno-solutionist" view that technological solutions to human problems are *always* appropriate or superior to other modes of intervention (Morozov, 2013). When the goal is to unlock positive human futures, longer term institutional reform and collective action remain important levers of change.

[5] We use the term "technologists" to denote the individual and institutional actors who exercise decision-making authority over the design, development, funding, management or deployment of a technology. This definition moves the focus beyond specific occupational categories (such as "computer scientists" or "engineers") and toward operational agency within the innovation ecosystem (Winner, 2020). It includes the architects of a technology alongside the organizational leaders, investors, and policymakers who shape its real-world impact.

Turning to the question of duty-bearers and the role of technologists, there are two major schools of thought. According to the *institutionalist* understanding of human rights, states are the first port of call when it comes to human rights, and they have stringent duties to protect and serve those who reside within their borders in certain ways (Rawls, 2001). Yet, when states fail or are unable to perform these duties, the responsibility for ensuring that these needs are met may transfer to other actors, including, for example, wealthier nations or whoever has the capacity to fulfill these duties (D. Miller, 2007). In situations of this kind, "waves of duty" tend to fan outward and arrive at the feet of those who have the greatest capacity to assist (Waldron, 1989). When advanced technologies provide the right kind of capacity, human rights claims will attach to the custodians of these technologies on the grounds that they are most able to provide the required form of assistance.

According to an *interactionist* understanding of human rights, things look quite different: the duties associated with human rights apply to everyone directly, potentially including those who develop new technologies. However, the duties generated by positive human rights are, to use a Kantian term, *imperfect* duties. This means that they are stringent, but that individuals, corporations, and nations have discretion over how to discharge these duties. Typically, two concerns should guide duty-bearers in this situation. First, parties should aim to do their fair share (Murphy, 2000). If, as suggested, technologists are well placed to contribute relative to other actors, the duty to do so may reside first and foremost with them. Second, in situations of partial compliance, where other actors are manifestly failing to help, there can be further duties to help "pick up the slack" (Karnein, 2014; Stemplowska, 2016).[6] Thus, even if there are better ways to meet human rights needs in theory, the ability of technology to circumvent institutional barriers that impede other modes of assistance means that those controlling the technology may inherit further responsibilities to support human rights goals.[7]

Finally, the claims generated by human rights, both to technology and other forms of assistance, only exist when human rights are not being respected or fulfilled. The existence of such claims is therefore likely to be hugely consequential in a world such as our own, marked by poverty, institutional failure, and widespread inaction (Gabriel, 2013). However, there is also an upper bound on what human rights-based arguments require: once they are met, there will be no further claim to benefit from technology on this basis. This finding differs subtly from the Central Claim, which is open-ended and holds, by hypothesis, even for worlds where there are no human rights shortfalls or violations at all. The Human Rights and Essential Goods argument therefore provides strong but conditional support for the Central Claim. For arguments that are more general and fully open-ended, we need to turn our attention elsewhere.

## 2. The Global Benefit and Squandered Potential Argument

Science fiction literature has provided us with many visions of worlds where the affluent and the destitute live "cheek by jowl" despite the existence of technological abundance. In these scenarios, gleaming towers and lavish tastes often find expression "up there," while those down below can barely afford the basic amenities needed to get by. In some versions of this story, those with access to advanced technologies also live significantly longer lives or enjoy off-world travel that is unavailable to the rest of the population. Suffice to say, much has gone wrong in these societies, not least the level of coercion and oppression typically required to maintain control over such a volatile situation. Yet, placing these political concerns to one side, we are also prompted to ask a deeper moral question: Why—given such technological wealth—are the benefits of technology not shared more widely? Indeed, this question needs to be answered even when those at the bottom are not enduring extreme hardship. In both cases, the failure to use advanced technology for the wider good results in a profound waste of human

[6] The question of how responsibility should be allocated in situations of partial compliance is a central concern of *non-ideal* theory (Rawls, 1999; Sen, 2009). While philosophers have often been concerned with identifying the appropriate ideal for a society, we also need to consider and perhaps prioritize philosophy for a world such as our own, where just institutions often do not exist and where there is widespread non-compliance with moral requirements (Mills, 2005; Murphy, 2000).

[7] This argument is also relevant in the context of claims about technology and injustice. It receives a more detailed treatment in Section 5 of this paper.

potential—to the loss of the flourishing and happiness that might otherwise have been.

As these scenarios make clear, morality is about more than just rights and duties, understood as a set of minimal human rights claims. There are also other kinds of considerations that determine what it means to act well in an interconnected world such as our own. Some of these considerations are particular to specific contexts and relationships, but others, such as those involving *beneficence*, are likely to be global in scope. One of the clearest accounts of beneficence is provided by act consequentialism, which is a moral theory that holds that we are all required to promote the best overall state of affairs through our individual actions—and that it is wrong not to do so (Scheffler, 1994). According to act consequentialists, this requirement applies to everyone, including those who invent and distribute new technologies. Indeed, from a consequentialist perspective, the actions of technologists will often be particularly important, given the amount of power they have to shape the future (Gabriel and Ghazavi, 2022; Jonas, 1984), including how benefits and burdens are distributed on a global basis.

Act consequentialism supports the Central Claim in a relatively direct way. If advanced technologies are transformational and beneficial, and if there is a general obligation to promote the greatest good for the greatest number, it follows that there is a moral obligation to ensure these technologies are designed, distributed, and deployed in ways that achieve this outcome. Indeed, scientists or entrepreneurs whose actions significantly deviate from this path wrongly allow preventable suffering to continue and squander significant human potential. Looking forward, it is possible to imagine worlds where technologies like AI *could* grant everyone a high standard of living and promote sustainable growth, but where these gains are captured by a small subset of actors, leaving many in a condition of relative material scarcity. Given the availability of significantly better states of affairs, the consequentialist holds that the choices contributing to this negative outcome represent a serious moral failing.

However, this argument quickly encounters an important challenge, namely, the contested nature of act consequentialism as the correct account of what morality requires. Among other things, opponents have argued that this interpretation of morality is excessively demanding (Railton, 1984; Scheffler, 1994; Smart and Williams, 1973), that the focus on promoting impartially better states of affairs licenses impermissible trade-offs between people (Rawls, 1999; Scanlon, 1998), and that it fails to adequately accommodate fundamental rights (Gewirth, 1982). Grounding the Central Claim in this way is therefore problematic if we want it to be acceptable to a wide audience of people. Fortunately, the current argument does not need to be framed in such stringent terms, nor so tightly bound to a purely consequentialist moral philosophy (Cullity, 2006; Singer, 1972). In fact, many less stringent accounts of beneficence support similar conclusions about advanced technologies. For example, regardless of our specific beliefs about morality, many of us would agree that if we can achieve something extraordinarily good, at little or minor cost to ourselves, we ought morally to do that thing. Yet this "weaker" account of beneficence leads to a similar overall conclusion with regard to the Central Claim: by distributing the benefits of an advanced technology widely, those who develop it are often in a position to do something very significant for humanity at little or minor cost, hence it would be wrong not to do so.

The latter argument does not rest on act consequentialism being the correct moral theory. Instead, it derives support for the Central Claim from a more moderate thesis that can be supported by various ethical perspectives. It also has several other advantages when compared with the Human Rights and Essential Goods argument. Rather than needing technology to be instrumentally important for achieving specific goods, such as health or physical security, which are protected by human rights, the global benefit argument is fully general. It applies to *any* potential benefit that advanced technologies may create. Beneficence-based arguments also provide robust support for the Central Claim in other ways: unlike human rights arguments the claims they generate are not "capped," and they do not rest upon the global situation being morally urgent or one in which there is acute unmet material need, although these features may make claims of beneficence more compelling.

Nonetheless, there are further objections to the global benefit argument that need to be addressed. First, the revised, weaker version of this argument makes use of a "little or minor cost" proviso, but its applicability across a range of contexts may be contested. For example, the costs involved in creating transformational GPTs such as advanced AI may not be minimal in *absolute terms*, as such artifacts are often very resource intensive to build. Efforts to develop the technology may also represent a huge expenditure of *personal resources* and *effort* on the part of those involved in this endeavor. Taken together, the creation of advanced technologies often requires extraordinary effort, coordination, and resources. As a result, even if we accept that there are general duties of beneficence, we may nonetheless doubt that they hold for technologies that are expensive and difficult to build. Developing this perspective further, one might acknowledge that it would be morally good for those who develop advanced technologies to use them in ways that benefit humanity, but entertain wider doubts about beneficence and whether these things are truly morally required (Herman, 1993; Narveson, 2003).

These are valid queries, but in each case a response is also available. First, with regard to the minimal cost proviso, the most morally relevant conception of "cost" understands this notion in *relation* to the potential gains (Ashford, 2003) that can be unlocked by an action. On this view, it seems likely that relative to the potentially vast global gains enabled by the wide distribution of benefits from advanced technologies, most of the factors discussed above remain relatively minor. This is especially true when we consider the magnitude of the benefit that advanced technologies are positioned to unlock over longer periods of time. Nonetheless, there are other interpretations of morality (R. D. Miller, 2004) that hold that the absolute cost parties incur *is* always relevant when determining the content of duties owed to others.[8] This view and its implications warrant further scrutiny (Cullity, 2006; Gabriel, 2017) (see Section 6). Deeper investigation of the empirical and moral costs involved in building advanced technologies may also prove illuminating. However, at the limit, the assumption that morality cannot impose substantial costs upon people, in general, is mistaken (Kagan, 1989; O'Neill, 2000) – a finding that carries over into the realm of technology. Fortunately, well-designed institutional frameworks can potentially mitigate some of these trade-offs, unlocking widespread benefit while simultaneously securing returns for inventors. Moreover, the strength of the Central Claim does not turn upon the moral significance of beneficence understood in isolation. There are a number of other supporting arguments to which we now turn.

## 3. The Global Circumstances of Birth Argument

In the Story of Er, as recounted by Plato, a comatose soldier lies inert on the battlefield and temporarily proceeds to the afterlife. There he sees a parade of souls: those moving on toward the next world and those poised to return to this one. In the latter case, lots are drawn to decide who will be born where, with what traits, and in what form. Yet the parable does not end there. Plato concludes that, with sufficient moral wisdom, even those who fare poorly in the lottery should be able to turn things around and lead a fulfilling life. Unfortunately, in the world as it is now, it is far from clear that this is the case. Many children do not make it to adulthood due to malnutrition and disease, and among those who do make it, fortunes tend to be constrained in a number of ways (Galasso and Wagstaff, 2019; Strauss and Thomas, 1998). Indeed, after laying out the facts about global inequality, the data scientist and sociologist Max Roser concludes that "our own hard work and life choices matter... much less than the one big thing over which we have no control: where and when we are born. This single, utterly random, factor largely determines the conditions in which we live our lives" (2019).[9]

[8] For example, the philosopher Richard Miller defends the idea that the evaluation of moral principles governing assistance should be subject to a "non-comparative" standard for reasonable rejection (R. D. Miller, 2004). On this account, what matters is how great an absolute cost moral principles place upon each party, in expectation, without such cost being offset by greater gain to others. For arguments against the idea that this generates wide-ranging permission to do as one pleases, see Gabriel (2017).

With these factors in mind, it strikes many people as self-evident that the world we live in is not a fair place. However, there is less agreement about the specific way in which the world is *unfair*. The simplest and perhaps most plausible interpretation of this view begins with the observation that luck—and specifically a person's place of birth—has an astonishing impact on a person's overall life chances (Milanovic, 2015). For example, a person born in sub-Saharan Africa is *fifteen* times more likely to die before their fifth birthday than someone born in a high-income country (UNICEF, 2025). They are also significantly less likely to complete secondary school education, something that significantly influences their opportunities later in life (Psacharopoulos and Patrinos, 2018; UNICEF, 2025). This situation is problematic when combined with a common intuition about justice: that it is bad (because it is unjust and unfair) that some people should be worse off than others through no fault of their own (Temkin, 1993: 13).

If this overall picture of things is correct, and the current global situation is unfair in the way that has been described, this has a range of moral consequences both for advanced technologies and more widely. Importantly, unfairness sometimes generates responsibilities for others to help remedy the situation (Caney, 2005; Lippert-Rasmussen, 2016; Tan, 2004). Applied to the present case, it suggests that those who enjoy abundant good fortune have a moral responsibility to try and improve the situation for those who do not—and, in doing so, to help produce a fairer state of global affairs. This argument, which is sometimes referred to as global "luck egalitarianism" (because of its focus on mitigating the impact of bad luck on a person's life chances), points toward the existence of responsibilities that encompass a wide range of actors.[10] On *institutional* accounts, states are the primary bearers of this responsibility and should try to discharge it—most likely via the transfer of resources and the reform of global institutions in ways that support fairer outcomes (Beitz, 1999; Pogge, 2002). On *interactionist* accounts, the duty-bearers are a still larger group of parties that includes individuals, companies, and non-state organizations (Caney, 2005). Yet, on both views, moral responsibilities arising from the need to address inequality bear upon the validity of the Central Claim in important ways.

[9] The inequality introduced by variation in geographical location is typically much larger than the difference to their life that a person can make on their own. For example, in a place where gross domestic product per capita is less than $1,000, it is almost impossible for a person to achieve what would be an average income in a rich country. On this point Roser notes that, "where you live isn't just more important than all your other characteristics, it's more important than everything else put together" (Roser, 2019; Roser, 2021).

[10] Relatedly, it is interesting to note that luck often plays a significant role in the process of technological invention and discovery. We have already stated that those who make such breakthroughs may be entitled to a return for their effort. But luck also plays a role in scientific discovery— a fact which is often overlooked—as the discovery of penicillin (in a stray Petri dish) or X-rays (via the glowing of a fluorescent screen) can attest to. Even in organized scientific endeavors, it is impossible to know ahead of time which avenues of investigation will bear fruit and which will not. This point further underscores the role played by collective distributed effort in scientific discovery (see Section 4 for related discussion).

Crucially, as we have seen, access to advanced GPTs is a major driver of growth and prosperity at the international level (Bresnahan and Trajtenberg, 1995; Krugman, 1979). It also has a profound effect on the lives of people in countries that are able to benefit from it, creating opportunities and wealth that manifest across a range of areas. Conversely, the uneven distribution of technology (and its benefits) helps to further entrench global inequality, perpetuating this inequality in life chances over time (Acemoglu, 2002). The redistribution of technology and its benefits therefore represents an important lever that parties can use to support fairer global outcomes (Huang and Manning, 2025; Dennis et al., 2025). Indeed, the Global Circumstances of Birth argument holds that the responsibility to use technology in this way is a by-product of responsibility that we all have (either directly or indirectly) to try to give people a fair chance in life.

What kind of support does this argument provide to the Central Claim? First, the argument is global in scope: it applies to everyone and does not turn on people being very badly off. Second, it is general as it applies to all advanced technologies rather than to specific technological artifacts. Third, it generates claims on technology that are not capped. Fairness-based arguments exist whenever there is unchosen inequality in life prospects. However, they become stronger in a more unequal world, as the impact of luck on a person's life chances begins to dominate other considerations. Fourth, the argument is quite minimal in terms of the empirical and normative assumptions it makes. Unlike the arguments considered later in this paper (Sections 4 and 5), the Global Circumstances of Birth argument says little about what caused global inequality to come about or why it continues to exist today. It is therefore unlikely to be refuted on causal or empirical grounds. It asks us only to acknowledge that people have radically unequal life chances, that this is unfair, and that the way in which the benefits of advanced technology are distributed intersects with these factors in normatively significant ways.

One objection to the Central Claim, when supported in this way, focuses on advanced technology and how relevant it is for the kind of task set out. We have already noted that the responsibility to address injustice may require a *range* of actions if the situation is to be satisfactorily addressed. So, why focus on advanced technology rather than other means? Might it not be better just to transfer resources that are by-products of the wealth created by technology when it comes to equalizing global chances? These questions point toward what we might refer to as uncertainty about the *currency* of global egalitarian justice. Why is the claim to benefit from technology and not to other things?

We have already encountered the seeds of an answer. First, although other mechanisms are important, we believe that decisions about advanced technologies represent an especially consequential lever to pull (Cockburn, 2018). If we look at other transformational GPTs (such as the steam engine, electricity, or the Internet), access to them, and the economic benefits they unlocked, helped shape global opportunities in a thoroughgoing way (van Dijk, 2005; Lucas, 2002; Mokyr, 1990). Effects of a similar or greater magnitude may hold for advanced AI (Korinek and Stiglitz, 2021). Second, efforts to distribute the benefits of technology are likely to be an effective way of achieving fairer global outcomes, even if they are not sufficient for this purpose. This is because the way technology is integrated into social and economic institutions shapes what opportunities people have available to them further down the line (Gabriel, 2022; Huang and Manning, 2025). Third, we live in a *non-ideal* world, meaning that we do not live in a world where other people always do their part in fulfilling shared moral commitments (Murphy, 2000). Thus, even if global economic resource transfer could help mitigate pernicious effects arising from the lottery of birth, it is not happening. As in the case of certain human rights commitments (Section 1), these responsibilities may increasingly "carry over" to those who have the capacity to act. When we know that other things will not be done, or that we ourselves are failing to take other actions geared toward discharging these responsibilities, it becomes even more important that technological levers are pulled to support fairer outcomes.[11]

Nonetheless, as with the argument from beneficence, elements of this fairness-based argument are philosophically contested. In particular, while many people share the intuition that it is unfair that the lottery of birth determines a person's life chances, there is less agreement that this fact by itself generates *responsibilities* to address this situation. Doubts are especially likely to be aired among those who adhere to a strong understanding of moral responsibilities—according to which the related claims can be coercively enforced by third parties—or for those who believe that duties only arise when we have wronged people or in certain institutional contexts (Nozick, 1974). Alternatively, as with arguments from beneficence, people may doubt that these moral claims are weighty enough to ground a responsibility to act.

At this point, advocates of fairness-based responsibilities may simply dig in and hold the line. They will note that, other things being equal, there are weighty moral reasons to support fairer outcomes over less fair outcomes—and that these reasons support the Central Claim. Indeed, in ordinary moral life considerations of fairness are widely held to be significant and to perform an important moral function in guiding behavior (Graham et al., 2013). However, the discussion does not need to end here. Crucially, we do not have to be "monists" about the grounds of justice, arguing that these responsibilities arise *only* as a by-product of the lottery of birth. In fact, pluralism—the view that both institutional and noninstitutional factors can ground these responsibilities—is more plausible (Cohen and Sabel, 2006; Risse, 2012). In Section 5, we argue that considerations of justice support the Central Claim because of factors tied to the character and operation of the world economy.

[11] On a deeper level, one might also argue that access to technology is *constitutive* of fairer outcomes insofar as it limits forms of relational inequality (Anderson, 1999) and off-sets the risk of domination (Pettit, 1997). Indeed, theorists since Adam Smith (1776) have long noted the role that access to certain goods—such as a linen shirt and leather shoes in 18th Century Britain—play in allowing citizens to interact on a level playing field and "look each other in the eye" (Sen, 1983; Pettit, 1997). Access to technological artefacts plausibly play a similar role in the world today. The idea that the dispersal of technology can counteract trends that would otherwise risk domination has been explored by Buterin (2023) and others.

## 4. The Global Tree of Knowledge Argument

Modern science and technology rest on the global tree of knowledge, an immense intergenerational stock of ideas, theories, data, and methods accumulated over millennia (Beinhocker, 2007; Muthukrishna and Henrich, 2016; Poskett, 2022). Every new scientific discovery or technological invention is a product of this *common inheritance* in the sense that it fundamentally builds upon earlier knowledge to contribute something new (Arthur, 2009). This process can involve extending existing work, refining it, or directly challenging and superseding it. For instance, Albert Einstein's theory of general relativity was built upon the foundational laws of motion and universal gravitation established by Isaac Newton centuries earlier, providing a new, more comprehensive theory that explained phenomena Newton's laws could not, such as the precise orbit of Mercury. This process continues today with the AI revolution, which is in many respects the physical instantiation of binary logic and algorithms proposed by mathematicians like Leibniz and Boole centuries ago, turning their abstract symbolic systems into the infrastructure of the modern world. The tree of knowledge is a *necessary background condition* for the creation of advanced technologies, including AI. Indeed, even with the resources, labor, infrastructures, and networks that today comprise the AI ecosystem, the development of large language models, deep learning systems, and AI agents (Kasirzadeh and Gabriel, 2025) would not be possible without drawing from a huge well of existing scientific and technological knowledge. These dependencies have significant implications for the distribution of technological benefits and the validity of the Central Claim.

To understand why, it is helpful to distinguish between two types of intertwined branches that comprise the global tree of knowledge. On the one hand, there is the *historical stock* of knowledge: the accumulated results of past inquiry and practice, from prehistoric toolmaking, to early modern mathematics and the twentieth-century digital computer revolution. On the other hand, there is the *current contributions* ecosystem, which is made up of a vast range of interlocking actors, efforts, and flows that enable the distributed, ongoing production of fresh insights by researchers, engineers, and practitioners today. Key ingredients for this modern research ecosystem include education infrastructure, markets, communication systems, a healthy labor force, stability, and the rule of law, all of which support the creation of new knowledge in important ways. These branches together form a valuable *public good*—in the form of a shared and evolving body of knowledge—that no single person, firm, or nation would be able to create in isolation, regardless of their wealth, effort, or intellectual capabilities (Coe and Helpman, 1995; Romer, 1994).[12]

The existence of this vast body of accumulated work, geared toward human advancement on the micro and macro levels, and its importance for contemporary technological innovation and production, supports the Central Claim in two ways. The first argument centers upon the nature of the scientific undertaking as a *sustained intergenerational practice* geared toward a common purpose, aptly characterized by Enlightenment thinker Francis Bacon (1605) as the development of a storehouse of knowledge "for the relief of man's estate." According to this view, those who engage in knowledge consumption and production are part of an ongoing enterprise, sustained by certain norms and goals that are reciprocally upheld, to reap the benefits of a collectively beneficial scientific practice over time. Particularly relevant to the Central Claim is the norm of *communalism* (Merton, 1973), which dictates that people who draw out of the pool of knowledge should in turn contribute to it, thereby supporting the wider cascade of beneficial innovation.[13]

At a minimum, this means that ideas that have the potential to benefit humanity should be shared. Yet, in the present case, which centers on transformational GPTs, there is reason to think that such measures will not always be appropriate or go far enough. To begin with, if the technology is dual use in nature—bringing with it significant potential both for benefit and harm—certain restrictions on knowledge sharing may be necessary (Miller and Selgelid, 2007; Shevlane, 2023). In these cases, there is an opportunity to still make good on the promise of science by delivering benefit in other ways—such as through tools or services that directly improve people's lives, without generating attendant risk. Second, even when this is not the case, and knowledge can safely circulate freely, it is not clear that collective duties to share scientific benefit are best or *fully* discharged simply by sharing research findings. A richer, more sustained, contribution to the practice of science can be made by reinvesting its material returns more widely in the education, health, and well-being of people who will continue to support this endeavor, either directly or indirectly, moving forward.[14]

[12] Public goods have a number of important properties. They are non-rivalrous in the sense that if one person uses it, this does not deplete the stock available for others (Samuelson, 1954). They are also non-excludable in the sense that it is hard, or impossible, to prevent other people from benefiting from these goods once they exist—even if they do not pay for them (Cornes, 1996).

[13] Commenting upon this interdependence—and the norms of reciprocity that sustain it—Albert Einstein wrote that: "A hundred times every day I remind myself that my inner and outer life are based on the labors of other men, living and dead, and that I must exert myself in order to give in the same measure as I have received and am still receiving..." (1935).

In the context of the Central Claim, it seems clear that those who push forward technological advances while, in the words of Sir Isaac Newton, "standing on the shoulders of giants" should give back to the community and not damage or degrade the common stock of knowledge (Huang and Siddarth, 2023). Yet critics may wonder whether this return also includes delivering the kind of broad-based material benefit identified by the Central Claim. After all, the philosopher Robert Nozick has argued that those who passively benefit from non-excludable public goods are not typically obliged to pay for their sustenance, or make wider community contributions, even if such payment is later called for (Nozick, 1974, 93–95).[15] Yet, in the present case, greater nuance is called for. To begin with, the historical stock of knowledge is the product of significant collective effort. There is a huge publicly funded infrastructure—including the interlocking efforts of many people and institutions—that supports the continued generation, open distribution, and accessibility of knowledge at a global scale. The scale of this effort and of the benefit unlocks calls for recognition and return (Arneson, 1982). More importantly, in most cases, the use of this knowledge by technology developers is not passive in the aforementioned sense.[16] Rather, the development of modern AI systems, for example, involves a huge industrial scale effort to use public knowledge to train and advance state-of-the-art systems (Crawford, 2021). Under these conditions, where commercial value is actively generated from the proceeds of a joint undertaking, the Tree of Knowledge argument holds that there is a responsibility to ensure broad material benefit.

Finally, an objection might nonetheless be raised that, unlike those who draw upon other public resources, the people who build advanced technologies today never chose to actively become part of a global scientific practice: they were born into a world that contains vast troves of knowledge and the practices that sustain it. As with citizens born into states, they simply inherited these riches—and doubts have long been held about the state's authority to enforce the law under such conditions (Simmons, 1999). Yet, here too the details matter. Crucially, while doubts surround the coercive enforcement of claims in the absence of explicit consent, there is much less disagreement about what reciprocity or fair treatment requires in these contexts (Arneson, 1982; Klosko, 2008). By becoming part of a practice from which you benefit significantly and that is also geared toward collective well-being—as in the case of modern scientific practice—one becomes part of a moral enterprise that is capable of generating a moral responsibility to, in turn, make a fair contribution to this practice.

---

[14] The importance of reinvesting deeply and across the entire innovation ecosystem is underscored by the phenomenon of "lost Einsteins"—people who would have gone on to produce high-impact inventions if they had become inventors, but who grew up without the means to do so (Bell, 2018). Further research has suggested that this is a sizable group of people worldwide (Agarwal, 2023).

[15] Nozick's argument (1974) is specifically about whether one would thereby incur an obligation that others could legitimately enforce using coercion. It remains possible—and one might even think likely—that considerations of fairness create a responsibility to contribute, even though it would be wrong for others to subsequently try and coercively enforce this outcome (Lyons, 1993; Simmons, 2020).

[16] Passive benefit is value that is received from a public good automatically. Active benefit is generated when a person or organization intentionally chooses to engage with the good or utility in question in order to unlock value (Simmons, 2020).

The second Tree of Knowledge argument complements the first and focuses on the importance of ongoing, broad-based contributions to the *contemporary knowledge production ecosystem*. On this view, technologists build upon not only the work of older generations of scientists and innovators but also contemporary practice, which they are also enmeshed in and beneficiaries of. Almost everyone contributes, either directly or indirectly, to this practice and it is global in scope. To get a clearer handle on this point, we should ask: What does it take to achieve a breakthrough on the scale of modern AI systems? A number of major public inputs have already been listed. They include: a healthy and highly educated workforce, elite educational institutions and scientific laboratories, data from the digital commons, physical infrastructure, including energy and transport grids on a massive scale, and a global system of economic trade and finance. The technologist Kate Crawford notes that, when viewed in this way, AI is, among other things, "a manifestation of highly organized capital backed by vast systems of extraction and logistics, with supply chains that wrap around the entire planet" (2021). We argue in Section 5 that the structure of the world economy may itself create moral claims that bear upon the distribution of advanced technology, such as AI, independently of how that particular technology is produced. However, the second Tree of Knowledge argument, considered here, focuses specifically on the inputs to technological invention. It holds that all who contribute to this venture, even indirectly, are owed something in return.

This notion can best be understood in relation to the idea of fair reciprocity and to the notion that everyone who contributes to an effort should get something back from it. Put simply, people should not be shortchanged.[17] Distributing the technological dividend widely, by honoring the Central Claim, helps to ensure that this is the case.

[17] In the words of the philosopher John Broome, "equal claims require equal satisfaction, stronger claims require more satisfaction than weaker ones, and also, very importantly [...] weaker claims require some satisfaction" (1990, 95). So, those who contribute something significant should not receive little in return, and those who contribute something should not receive nothing in return. In both cases, they are owed something more.

Importantly, while those who produce technologies typically do already pay back in a number of ways—via the payment of salaries, taxes, and the purchase of resources, goods, and services—it is not clear that people are widely and fairly reimbursed for their contributions. This becomes clear when we consider, for example, miners in far-flung countries digging for rare earth minerals, educators around the world who support the global stock of knowledge, people whose job is to support the operation of basic infrastructure such as transport systems, and everyone who has contributed to the public revenues needed to support this undertaking (Bratton, 2026; Crawford, 2021; Gray and Suri, 2019; Lanier and Weyl, 2018). According to the contemporary knowledge production argument, delivering on the Central Claim is the best way *in expectation* to ensure that those whose efforts hold up the global tree of knowledge continue to be adequately supported and sustained.

Turning to objections, one might wonder whether this argument really supports an entitlement to benefit that is fully universal. Is it true that *everyone* contributes to global knowledge production and to the creation of advanced technological artifacts such as modern AI systems? After all, these breakthroughs are often geographically concentrated, whereas the universality of the Central Claim encompasses all people, including fishers, farmers, young children, and the inhabitants of remote island communities.

This question can be answered on two levels. One approach takes us into deeper philosophical waters and concedes that the scope of claims may not be fully universal unless it is also the case that some of the natural resources used in technology production are owned in common by all of humankind. Such arguments have strong intellectual lineage, including in Catholic social teaching (Pope Leo XIV, 2026), in Thomas Paine's (1797) notion that earth in its uncultivated state is the common property of all, and in the contemporary philosophy of theorists such as Hillel Steiner (1994) and Peter Vallentyne (2000) who defend the notion of a global resource trust. What follows from common resource ownership, may well be an entitlement to receive a return on resources that are consumed or used by others. However, even without this further argument, a strong presumptive case in favor of universal benefit can often be made. Practically speaking, given the impossibility of parsing the nature of global contemporary contributions to the tree of knowledge on a granular level, and the high confidence that such contributions have been both made and proven foundational to the success of this effort (Poskett, 2022), the best practical approach to discharging the relevant moral duties is to honor the promise of the Central Claim and ensure that benefits are widely distributed and accrue to all. In the case of children, such a rationale may need to be extended to encompass investment in the continued health of the knowledge ecosystem, potentially with reference to the first Tree of Knowledge argument, which centers upon historical contributions and the claims they generate.

Clearly, people may still disagree about *how much* people are owed and about whether this dividend should be divided evenly. Nonetheless, on a more basic level, this argument also suggests that in an imperfect world, duties of fair reciprocity support the Central Claim.

## 5. The Global Economic Institutions Argument

We have already noted that the production of new technologies tends to rest upon a large network of global economic activity. According to the arguments in the previous section, widespread contributions to technological breakthroughs support a claim to benefit from the technologies that are the product of this process. However, the existence of a global economy—made up of interlocking rules, laws, and agreements, underwritten by coercive sanctions—also has wider implications for global morality. Indeed, many theorists have argued that it serves as the basis for international principles of distributive justice (Beitz, 1975; Caney, 2005; Tan, 2004). If these arguments are correct, the Central Claim may also be supported by broader demands of justice that are valid *independently* of specific claims about how a given technology is created. The Global Economic Institutions argument, as outlined below, provides the final freestanding foundation for the Central Claim surveyed in this paper.

The crux of the Global Economic Institutions argument turns on a common understanding of the role that the principles of justice are intended to serve. While justice is sometimes taken to refer to a basic moral property, akin to the notion of fairness discussed by luck egalitarians in Section 3, it also often refers to something more than this: to a system of principles that determine how a society's major institutions should operate, including the distribution of wealth, rights, and opportunities, in such a way that it can command the full support of its members (Rawls, 1999; Rawls, 2005). Ultimately, the goal of such principles is to support the fair division of advantages that arise from social cooperation and to justify state power when it is used in service of these ends, including via the law and its coercive enforcement. The need to justify these practices, in turn, generates a strong set of demands: the state must protect individual rights, support equality of opportunity, and promote a fair distribution of resources (Rawls, 1999). Falling short of this, people would not be able to consent to political authority in the right way and could be said to experience domination.

The key insight, for the Global Economic Institutions argument, is that many of these processes are no longer bound exclusively to the territorial nation-state in the modern age. Instead, a number of theorists have argued that there are *global practices* that also have the relevant properties and that call for similar kinds of justification (Beitz, 1999; Caney 2005; Cohen and Sabel 2006; Pogge, 2002; Tan, 2004; Young, 2011).

[18] Three features of the global economic system stand out. First, it involves *reciprocal cooperation* on a large scale via intricate networks of exchange and flows that make up global trade, capital markets, and the movement of people. Second, the system is *coercive* insofar as it is governed by a system of international institutions, treaties, and sanctions that determine the rules of the game and punish transgressors (Farrell, 2019). [19] Finally, the design of the global economic system has a *profound* and *nonvoluntary effect* on people's lives, as those who have been helped or harmed by economic globalization can attest. Life chances are significantly shaped by place of birth (see Section 3), but they are also significantly shaped by everything that happens after that point. These three properties ground further claims about distributive justice. Indeed, they suggest that we are either all already subject to a global system of interaction and exchange to which principles of justice readily apply, or that we are subject to a global system that ought to be regulated by principles of justice rather than simply being left to chance (Abizadeh, 2007).

When it comes to articulating principles of justice that apply to the world economy, a range of positions have been put forward. Theorists such as Simon Caney (2005) and Charles Beitz (1999) have argued that these principles should be the same as those that operate at the level of the nation-state, such as fair equality of opportunity and efforts to ensure that economic inequality is to the overall advantage of the worst off (Rawls, 1999). Alternatively, theorists such as Joshua Cohen and Charles Sabel (2006) and David Miller (2007) have argued for egalitarian or sufficientarian principles of global justice that are somewhat weaker than the principles that apply in domestic settings.[20] Yet what matters most for the Central Claim is that even on less demanding accounts of global justice, the current world economy falls far short of these standards.

From a distributive perspective, high levels of global material inequality, and the prevalence of hunger and stunting—which affect 650 million people and 150 million children, respectively—provide strong *prima facie* grounds for thinking existing global institutions are unjust. Scholars have also highlighted specific global practices that drive these outcomes (Chang, 2001; Pogge, 2002; Wenar, 2015). Prominent examples include asymmetric trade and investment rules that often shelter industries in high-income countries while destabilizing those of developing economies (Christensen, 2017; Woods 2006), the international resource privilege, which sanctions the unconstrained appropriation of natural resources by national political elites (Pogge, 2002; Tørsløv et al., 2023; Wenar, 2008), and limited extraterritorial liability, which grants transnational corporations impunity in various settings (Ruggie, 2013). These practices cause foreseeable hardship. Indeed, the philosopher Iris Marion Young (2011) argues that the global economy involves structural injustice in which existing "processes put large groups of persons under systematic threat of domination or deprivation of the means to develop and exercise their capacities" while "enabling others to dominate or enjoy vastly greater opportunities."

[18] Indeed, the most famous version of this view holds that principles of distributive justice apply only to the *basic structure of society* defined in terms of "the political constitution, the legally recognized forms of property, and the organization of the economy" (Rawls, 2005, 258).
[19] For example, while goods move relatively freely, the movement of people is heavily restricted (Abizadeh, 2007).
[20] Sufficientarianism, as defined by the philosopher Ingrid Robeyns (2017), is the view that, "distributive justice should be concerned with ensuring that no one falls below a certain minimal threshold, which can be either a poverty threshold or a threshold for living a minimally decent life."

The Global Economic Institutions argument supports the Central Claim because, in situations where background injustice exists, duties to remedy the situation are often widely distributed, extending to include all who are part of that system. In the present case, a common thought is that the responsibility to address global injustice resides first and foremost with wealthy nations or with those who benefit most from global supply chains. Direct measures to support fairer global practices could include reforms to the global trade regime, tighter regulation of extractive industries, and efforts to redistribute wealth on a global level. However, there is reason to think that technology and technologists have a central role to play. First, as documented in Section 4, they are beneficiaries of these practices and hence may have a special reason to help remedy this state of affairs (Butt, 2007; Young, 2011). Second, even when they are not the primary duty-bearers, the responsibility to address injustice extends outward—toward those who have the power and the capability to intervene effectively (D. Miller, 2007; Shue, 1988). Indeed, when states and other powerful actors do not take steps to help mitigate injustice at the global level, other parties can be called on to help remedy the situation (Karnein, 2014; Stemplowska, 2016). Third, technology is an unusually *powerful lever* when it comes to shaping developmental trajectories and living standards on the global stage (Krugman, 1979; Acemoglu and Johnson, 2023).

Economists and economic historians have long shown that the ability to access and harness the material fruits of technological innovation is a critical factor for economic development (Bresnahan and Trajtenberg, 1995; Krugman, 1979; Mokyr, 1990; Solow, 1957), with the industrial revolution and industrialization representing major contributors to differential living standards in the world today (Lucas, 2002; Pomeranz, 2009). Moving forward, similar forecasts now apply to the impact of advanced AI systems, with Korinek and Stiglitz (2021) arguing that "progress in artificial intelligence and related forms of automation technologies threatens to reverse the gains that developing countries and emerging markets have experienced from integrating into the world economy over the past half century, aggravating poverty and inequality." Yet, as Korinek and Stiglitz are keen to stress, none of these outcomes are preordained. Rather, they call for concerted efforts to ensure that potential benefits from AI are widely shared and directed toward broader development goals. In the context of the Global Economic Institutions argument, this body of work suggests that efforts to exclude decisions about the development and distribution of technology from the set of relevant moral options would be to ignore an important *site of justice* (Buchanan, 2011; Gabriel, 2022) at the global level. Neglecting this dimension would amount to approaching the problem with one hand tied behind our back.

Taken together, these arguments support the Central Claim by arguing that efforts to ensure that everyone can benefit materially from major technological advances are a way of meeting the wider demand for justice in the world economy. The Global Economic Institutions argument holds that, against a backdrop of unjust global practices, there are countervailing duties to ensure that benefit is shared widely and that the fruits of advanced technology are shared by all. As Iris Marion Young (2011) argues, all agents contributing to the ongoing operation of this structure, including citizens of wealthy countries, transnational corporations, and international bodies, share a responsibility to change it. Indeed, those in positions of greater power potentially hold a greater share of this responsibility. When the custodians of advanced technology have the capacity to mitigate or reverse the impact of unjust global practices on different populations, the Global Economic Institutions argument holds that there is a positive responsibility to use the benefits unlocked by advanced technology in this way.[21]

## 6. Open Questions and Implications

This paper has explored five moral arguments in support of the Central Claim: the proposition that the benefits of advanced technology belong to the world in the sense that everyone is entitled to benefit materially from their invention and use. While these arguments build upon different normative premises, they achieve a striking convergence—forming an overlapping consensus in support of the view that the benefits of advanced technology should accrue to all. We conclude by providing an overview of these arguments, discussing open questions, and providing a fuller picture of what the Central Claim may require of technologists in a global age.

The case for the Central Claim is multifaceted. First, the Human Rights and Essential Goods argument holds that universal access to beneficial technology is often a moral necessity. If human rights are universal and require key goods to be realized, and if advanced technology is now the primary means of providing those goods, the technology must itself be understood as something to which there is a valid claim. Second, the Global Benefit and Squandered Potential argument holds that failure to share the benefits of advanced technology would lead to significant preventable loss of human well-being and potential worldwide. Drawing upon a broadly consequentialist line of argument, it holds that if developers of advanced technologies have the power to create a significantly better state of affairs for large numbers of people, then they also have good moral reason to do so. This argument obtains even on weaker accounts of beneficence, given that the cost of providing assistance via technology is often small when compared to the benefit it would unlock. Third, the Global Circumstances of Birth argument holds that it is unfair that life prospects are so dramatically shaped by place of birth. Since advanced technologies are a significant driver of wealth and opportunity at a global level, technology should be used to help achieve fairer global outcomes. Fourth, the Global Tree of Knowledge argument holds that technological breakthroughs are best understood as a product of our collective human inheritance rather than as isolated or private achievements. Modern technology is the physical instantiation of an intergenerational stock of knowledge for which no single entity can claim exclusive credit. As technologists actively generate value from this shared venture, using everything from centuries-old logic to modern publicly funded infrastructure, they incur a debt of reciprocity to the global community. This debt is further deepened by the vast, contemporary ecosystem of global labor, data, and resources that makes technological innovation possible on a planetary scale. Finally, the Global Economic Institutions argument holds that the existence of a globally integrated economy, governed by coercive rules and trade agreements, generates its own set of moral requirements. The current system is affected by economic injustice, as asymmetric trade rules and resource exploitation which systematically deprive many people of the means to flourish. Since advanced technologies are a major driver of economic outcomes, they represent an important lever that must be pulled to counteract these institutional failures.

[21] The responsibilities of technologists may extend beyond the kind of material benefit which is the subject of the Central Claim. There may also be a proactive responsibility to build legal regimes and institutions, for technology and technological diffusion, that ensure these processes operate in a way that is consistent with principles of justice at a global level.

The arguments presented here are not intended to be exhaustive. There are further arguments, both in support and in opposition to the Central Claim, which have not been discussed—or were only touched upon briefly—due to limitations of space. Taking each side in turn, those who support the Central Claim could make further appeal to principles of:

(1) *Non-Domination*: widespread access to advanced technologies may be an important check on the ability of some parties to unjustly impose their will on others;
(2) *Historical Injustice*: access to the benefits of advanced technology may help rectify past wrongs, especially insofar as they continue to bear upon the economic development of nations today;

(3) *The Imposition of Risk*: the development of advanced technologies often brings with it exposure to risk that is widely dispersed, with compensation or payment potentially owed to those affected;
(4) *Human Solidarity*: the fact that we are all part of a common enterprise, progressing together through time, may ground further responsibilities to support one another and ensure that no one is involuntarily left behind, including in the realm of technology; and
(5) *Common Ownership of the Earth*: the common global ownership of natural resources (touched upon in Section 4) may generate a broad entitlement to returns from them, including when they are used to build advanced technologies.

Conversely, opponents of the Central Claim might argue that those who invent or develop technologies acquire a *special right* to reap the rewards of their innovation, including any financial returns that attach to it and the ability to insist on exclusive use. This view finds partial support via arguments already articulated in the paper. For example, we noted in our discussion of human rights (Section 1), that Article 27 also includes protections for "moral and material interests" arising from "any scientific, literary or artistic production of which he is the author." The Global Benefit and Squandered Potential argument (Section 2) takes into account the costs borne by technologists. The luck egalitarian perspective (Section 3) maintains that people are *not* responsible for unchosen misfortune. However, the flipside of this is that there may be a *positive entitlement* to accrue benefit when it is created through the voluntary and successful exercise of effort or skill. Finally, the contemporary Tree of Knowledge argument (Section 4) is sensitive to the level of contribution that different actors make—holding out the possibility that major contributions deserve larger returns. From a moral vantage point, all of these considerations are consistent with the Central Claim: it is possible for everyone to be owed something significant, and for some people to be owed more than others. What is not consistent with the Central Claim—and the position we reject in light of the arguments presented in this paper—is the moral claim to *exclusive* benefit from advanced technology.[22] Indeed, while there are both moral and instrumental reasons to ensure that those developing and producing advanced technologies receive a fair return for their effort and investment, the moral case for global benefit remains deeply compelling.

[22] Arguments supporting an exclusive right to benefit from scientific or technological discoveries are also difficult to ground on their own terms. Libertarian accounts of property ownership often assume a logical connection between self-ownership and the enforceable ownership of external objects (Nozick, 1974). However, significant distinctions exist between the two (Cohen, 1995; Fried, 2004). While self-ownership primarily concerns the individual, property rights are multilateral, imposing duties on others to refrain from using the thing in question (Waldron, 1990). The central challenge is therefore to explain how a unilateral act—such as labor, invention, or enclosure—creates a moral obligation for everyone else to respect an exclusive claim. This difficulty is compounded when other people hold independent claims in relation to the good (Sections 1–3) or when its production involves multiple parties (Sections 4–5).

Without more comprehensive investigation of each argument and the way it bears upon specific advanced technologies such as AI, arriving at an all-things-considered judgment about what morality requires in any given case remains a complex task. Nonetheless, the arguments developed in this paper provide an important starting point. We have established a robust *pro tanto* case for a universal entitlement to benefit from advanced technology, across a wide spectrum of cases, encompassing the majority of transformational GPTs. This finding has several significant practical ramifications.

First, for most advanced technologies, including AI, it is very likely that everyone has a claim to benefit materially from their invention and use. In this context, material benefit must be understood as access to goods or services that an individual is actually positioned to *experience*, rather than a theoretical opportunity contingent upon additional factors (van Dijk, 2020; Warschauer, 2004). Such benefits may be *direct*, as with the ability to use computers or access the Internet, or *indirect*, facilitated through intermediaries such as state or non-state actors. Furthermore, material benefit may be realized through the systemic redistribution of the financial returns generated by new technologies (O'Keefe et al., 2020). What is of primary importance is that the relevant moral claims and responsibilities are satisfactorily met, rather than the specific method of their achievement. When it comes to determining *how much* is required, the nature of the technology and the circumstances in which it is created shape the moral landscape in a number of ways. The force of the Central Claim is influenced by, for example, whether there are unmet human rights claims, how serious the underlying situation is, the size of the benefit this technology can unlock, the existence of other mechanisms for providing assistance, the various contributions that have been made to the process of technological innovation, considerations of background justice (or injustice) that connect different parties, and the action (or inaction) of other duty-bearers, such as states, including whether they are doing their fair share. Nonetheless, in almost all cases where the facts resemble those of the world today, something significant is owed to all.

Second, to meet this requirement, it is important to explore a wide set of institutional proposals and actions that could help unlock broad technological benefit. The Central Claim itself is largely *mechanism agnostic*. What it provides is a set of desiderata against which institutional proposals and actions ought to be assessed. Institutional arrangements that fail to account for these claims are presumptively unjust. Yet the question of how to achieve widespread benefit from technology is empirical, and it needs to be answered differently for different technologies, in a way that is sensitive to their value and to any risk that they create. Direct technology transfer or access is sometimes appropriate. This could involve giving the technology away for free or providing it in ways that are subsidized or cross-subsidized, thereby enabling widespread use. Alternatively, responsibilities may be satisfactorily discharged via broad market access, in situations where pricing constraints do not unduly limit uptake and use, or through other ordinary paths of technological diffusion (Coe, 1995; Ding, 2024; Jones and Summers, 2022; Nordhaus, 2004). In a further set of cases, we may need to set up new institutional partnerships or undertake further institutional innovation in order to address market failures or better harness economies of scale (Bloom et al., 2019; Buchanan, 2011; Kremer, 1998; Mazzucato, 2024; UNDP, 2025; World Bank, 2016). In each case, a key question is how these systems can unlock benefits while also incentivizing continued investment in new technologies, in a manner that supports long-term sustainable progress (Huang, 2021; Nordhaus, 1969). Indeed, the question of what constitutes a fair return to technologists is important not only from a moral perspective but also for ensuring that the scientific ecosystem continues to function well and can benefit everyone over time. Of course, intellectual property law is not a panacea: there are a number of well-documented cases where it has not performed this social function well (Chang, 2001; Drahos and Braithwaite, 2017; Maskus, 2000). However, well-calibrated legal regimes can potentially help to align private incentives and public goals across a range of cases.

In the context of ongoing conversations about how to unlock the potential benefits created by advanced AI, a variety of institutional mechanisms, including markets, global partnerships, new institutions, and wider efforts supporting redistribution, each have a role to play (Dennis et al. 2025; Ho et al., 2023; United Nations, 2024). For certain consumer-facing versions of AI, market mechanisms have already driven high rates of global uptake and use. However, there is reason to question whether the market, on its own, can meet the requirements of the Central Claim—especially in relation to a potentially transformational GPT. To begin with, the institutional preconditions for benefit are unevenly distributed (United Nations, 2024). In order for people to benefit from advanced AI, there needs to be greater investment in the infrastructure needed to run these systems in different geographies (Ding, 2024)—including via financing mechanisms and the provision of technical support. Additionally, the Central Claim bears both upon the *design* and deployment of advanced technologies. More specifically, it supports the co-development of new applications that unlock value for people living in parts of the world that have historically been under-served by market mechanisms (Mohamed, 2020; Peter and Devlin, 2025). The size of the gains that advanced AI could potentially unlock, provides further reason to consider what could be achieved through coordinated global action. If advanced AI really is an epoch-defining technology (Brynjolfsson, 2014), as we have countenanced, then there is good reason to believe that more can be done by partnering with states, international organizations and scientific institutions, than can be achieved by relying on markets alone (Kremer, 1998; Mazzucato, 2024). Finally, advanced AI may itself disrupt ordinary market dynamics or create forms of inequality that need to be managed via institutional efforts that promote fairer global outcomes (Korinek, 2021; Buchanan, 2011). To address these factors, a number of major studies have proposed the creation of new multilateral partnerships to widen access to computing power and talent (Dennis et al. 2025; United Nations, 2024), alongside programs designed to ensure that open AI models can be developed and refined in ways that render them appropriately valuable across diverse local contexts (UNDP 2025). Recently, the Secretary-General of the United Nations, António Guterres, has called for the development of a global fund for AI. More ambitiously still, a number of authors have proposed a "Windfall Trust" (O'Keefe et al., 2020)—envisaged as a trust for all living people, that would redirect a percentage of the returns from advanced AI towards globally beneficial outcomes.

Third, this investigation carries wider lessons for our understanding of technology and the role of technologists in a global age. The power of technology to shape global opportunities and states of affairs is greater now than at any previous point in history. Against this backdrop, it makes sense to focus more directly on the ethical responsibilities that accompany its creation and diffusion. Just as earlier moral frameworks, intended to govern interpersonal conduct in small communities, may no longer serve as an accurate guide to action in a world of globally networked dependencies (Scheffler, 2002), so the power of technology calls for a new kind of ethics—one that is commensurate with its impact on our own lives and on those of future generations (Jonas, 1984). On a more granular level, the transition to a world where technologists function as an important class of moral actor manifests in at least three ways. First, advanced technologies often bring with them unique capabilities, and hence the opportunity and responsibility to do things that could otherwise not be done *at all* (Jasanoff, 2016). Second, even when this is not the case, technologists often have a *unique* opportunity to unlock benefits in situations where others have chosen not to act or where alternative approaches appear to have stalled. Third, the development of advanced technology itself creates *special responsibilities* for technologists, understood as co-participants in a human practice from which they benefit and which is geared towards common ends. Given the high stakes involved in the development of advanced technologies—and their transformative potential—identifying correct principles to govern the benefits and risks they create will likely remain a central question for our times.

**Iason Gabriel**, Google DeepMind (iason@google.com)
**Atoosa Kasirzadeh**, Google DeepMind (atoosakz@google.com)

**Acknowledgements**: We would like to thank Leif Wenar, Henry Shue, Juri Viehoff, Johannes Himmelreich, Harvey Lederman, Alex Imas, Toby Ord, Sumaya Nur Adan, Kate Crawford, Varun Gauri, Vafa Ghazavi, Brian Christian, James Christensen, David Grewal, James Manyika, Juan Mateos-Garcia, Joseph Levine, Adam Bales, Seb Krier, Kevin Klyman, Julian Jacobs, Zhengdong Wang, Anna Koivuniemi, Ndidi Elue, Steph Hughes-Fitt, Alex Goldin, Shamil Chandaria, Thore Graepel and Shane Legg, for invaluable advice, comments and feedback on earlier drafts of this paper. I.G. would also like to thank Joshua Cohen and the faculty at the Department of Philosophy at UC Berkeley for the opportunity to work on this project as a Visiting Researcher in 2022.

## References

Abizadeh, A. 2007. “Coercion and the Grounds of Political Obligation: Theoretical Foundations and the Global Context.” *American Political Science Review* 101 (2): 333–42.

Acemoglu, D., 2002. “Directed Technical Change.” *The Review of Economic Studies*, 69 (4): 781-809.

Acemoglu, D., and S. Johnson. 2023. *Power and Progress: Our Thousand-Year Struggle Over Technology and Prosperity*. PublicAffairs.

Agarwal, R., I. Ganguli, P. Gaulé, and G. Smith, 2023. “Why US immigration matters for the global advancement of science.” *Research Policy*, *52* (1): 104659.

Aker, J.C. 2010. “Information From Markets Near and Far: Mobile Phones and Agricultural Markets in Niger.” *American Economic Journal: Applied Economics*, *2* (3):46-59.

Alesina, A. and D. Dollar. 2000. “Who Gives Foreign Aid to Whom and Why?” *Journal of Economic Growth*, *5* (1): 33-63.

Anderson, E.S. 1999. “What is the Point of Equality?” *Ethics* 109 (2): 287–337.

Arneson, R.J. 1982. “The principle of fairness and free-rider problems.” *Ethics*, *92* (4): 616-633.

Arthur, W.B. 2009. *The Nature of Technology: What It Is and How It Evolves*. Simon and Schuster.

Ashford, E., 2003. “The Demandingness of Scanlon’s Contractualism.” *Ethics*, *113* (2): 273-302.

Ashford, E., 2018. “Severe poverty as an unjust emergency.” *The Ethics of Giving: Philosophers’ Perspectives on Philanthropy*, Oxford University Press: pp.103-148.

Bacon, F. 1605. *The Advancement of Learning*. Dent.

Beduschi, A., 2022. “Harnessing the potential of artificial intelligence for humanitarian action: Opportunities and risks.” *International Review of the Red Cross*, 104 (919): 1149-1169.

Beinhocker, E.D. 2007. *The Origin of Wealth: Evolution, Complexity, and the Radical Remaking of Economics*. Random House.

Beitz, C. 1975. “Justice and International Relations.” *Philosophy & Public Affairs* 4 (4): 360–89.

Beitz, C. 1999. *Political Theory and International Relations.* Rev. ed. Princeton University Press.

Bell, A., R. Chetty, X. Jaravel, N. Petkova, and J. Van Reenen. 2019. “Who becomes an inventor in America? The importance of exposure to innovation.” *The Quarterly Journal of Economics*, *134* (2): 647-713.

Bloom, N., J. Van Reenen and H Williams. 2019. "A toolkit of policies to promote innovation." *Journal of Economic Perspectives*, 33 (3): 163-184.

Bratton, B. H. 2026. *The Stack: On Software and Sovereignty*. MIT Press.

Bresnahan, T. F., and M. Trajtenberg. 1995. "General Purpose Technologies: Engines of Growth?" *Journal of Econometrics* 65 (1): 83–108.

Brynjolfsson, E. and A. McAfee. 2014. *The Second Machine Age: Work Progress and Prosperity in a Time of Brilliant Technologies*. WW Norton & company.

Broome, J. (1990) "Fairness." In *Proceedings of the Aristotelian society*, vol. 91: 87-101

Buchanan, A., T. Cole, and R.O. Keohane, 2011. "Justice in the Diffusion of Innovation." *Journal of Political Philosophy,* 19 (3): 306-332.

Buterin, V. 2023. "My Techno-Optimism." *vitalik.eth.limo* (blog). November 27. https://vitalik.eth.limo/general/2023/11/27/techno_optimism.html.

Butt, D. 2007. "On Benefiting From Injustice." *Canadian Journal of Philosophy*, *37* (1): 129-152.

Camps-Valls, G., M.Á. Fernández-Torres, K.H. Cohrs, A. Höhl, A. Castelletti, A. Pacal, C. Robin, F. Martinuzzi, I. Papoutsis, I. Prapas and J. Pérez-Aracil. 2025. "Artificial Intelligence for Modeling and Understanding Extreme Weather and Climate Events." *Nature Communications*, *16*(1), p.1919.

Caney, S. 2005. *Justice Beyond Borders: A Global Political Theory*. Oxford University Press.

Chang, H. J. 2001. "Intellectual Property Rights and Economic Development: Historical Lessons and Emerging Issues." *Journal of Human Development*, 2 (2):287-309.

Christensen, J. 2017. *Trade Justice*. Oxford University Press.

Cockburn, I.M., R. Henderson, and S. Stern. 2018. "The Impact of Artificial Intelligence on Innovation: An Exploratory Analysis." In *The Economics of Artificial Intelligence: An agenda* (pp. 115-146). University of Chicago Press.

Coe, D. T. and E. Helpman. 1995. "International R&D Spillovers." *European Economic Review*, 39 (5): 859-887.

Cohen, J., and C. Sabel. 2006. "Extra Rempublicam Nulla Justitia?" *Philosophy & Public Affairs* 34 (2): 147–75.

Cohen, G.A.. 1995. *Self-ownership, freedom, and equality*. Cambridge University Press.

Cornes, R. and T. Sandler. 1996. *The Theory of Externalities, Public Goods, and Club Goods*. Cambridge University Press.

Crawford, K. 2021. *Atlas of AI: Power, Politics, and the Planetary Costs of Artificial Intelligence*. Yale University Press.

Crivelli, E., R. De Mooij and M. Keen. 2016. "Base Erosion, Profit Shifting and Developing Countries." *FinanzArchiv/Public Finance Analysis*: 268-301.

Cullity, G. 2006. *The Moral Demands of Affluence*. Oxford University Press.

Dennis, C., S. Manning, S. Clare, B. Wu, O. J. Effoduh, C. T. Okolo, L. Heim, and K. Klinova. 2025. *Options and Motivations for International AI Benefit Sharing.* Centre for the Governance of AI.

Ding, J. 2024. *Technology and the Rise of Great Powers: How Diffusion Shapes Economic Competition.* Princeton University Press.

Drahos, P., and J. Braithwaite. 2017. *Information Feudalism: Who Owns the Knowledge Economy*. Routledge.

Dreher, A., Lang, V. and Reinsberg, B., 2024. "Aid Effectiveness and Donor Motives." *World Development*, *176*: 106501.

Einstein, A. 1935. *The World as I See It*. London : John Lane.

Eloundou, T., S. Manning, P. Mishkin, and Daniel Rock. 2024 "GPTs are GPTs: Labor market impact potential of LLMs." *Science* 384, no. 6702: 1306-1308.

Farrell, H. and A. L. Newman. 2019. "Weaponized Interdependence: How Global Economic Networks Shape State Coercion." *International Security*, 44 (1): 42-79.

Fried, B.H. 2004. "Left‐Libertarianism: A Review Essay." *Philosophy & Public Affairs*, *32* (1):66-92.

Gabriel, I. 2013. *On Affluence and Poverty: Morality, Motivation and Practice in a Global Age*. Doctoral Dissertation, Oxford University, UK.

Gabriel, I. 2017. "The Problem with Yuppie Ethics." *Utilitas* 30 (1): 32–53.

Gabriel, I. 2022. "Toward a Theory of Justice for Artificial Intelligence." *Daedalus* 151 (2): 218–31.

Gabriel, I., and V. Ghazavi. 2022. "The challenge of value alignment." *The Oxford Handbook of Digital Ethics*. Oxford University Press: 327-340.

Galasso, E. and A. Wagstaff. 2019. "The Aggregate Income Losses from Childhood Stunting and the Returns to a Nutrition Intervention Aimed at Reducing Stunting." *Economics & Human Biology*, 34: 225-238.

Graham, J., J. Haidt, S. Koleva, M. Motyl, R. Iyer, S.P. Wojcik, and P.H. Ditto, 2013. "Moral Foundations Theory: The Pragmatic Validity of Moral Pluralism." In *Advances in experimental social psychology* (Vol. 47): 55-130. Academic Press.

Gray, M. L., and S. Suri. 2019. *Ghost Work: How to Stop Silicon Valley from Building a New Global Underclass*. Harper Business.

Gewirth, A. 1982. *Human Rights: Essays on Justification and Applications*. University of Chicago Press.

Griffin, J. 2008. *On Human Rights*. Oxford University Press.

Haykal, D. M. Goldust, H. Cartier and P. Treacy. 2025. "AI in Humanitarian Healthcare: A Game Changer for Crisis Response." *Frontiers in Artificial Intelligence*, 8: 1627773.

Herman, B. 1993. *The Practice of Moral Judgment*. Harvard University Press.

Ho, L., J. Barnhart, R. Trager, Y. Bengio, M. Brundage, A. Carnegie, R. Chowdhury, A. Dafoe, G. Hadfield, M. Levi, and D. Snidal. 2023. "International Institutions for Advanced AI." *arXiv preprint arXiv:2307.04699*.

Huang, S. and S. Manning. 2025. "Here's How to Share AI's Future Wealth." *Noema*, April 22. https://www.noemamag.com/heres-how-to-share-ais-future-wealth/.

Huang, S., and D. Siddarth. 2023. "Generative AI and the Digital Commons." *Preprint, arXiv March 20. https://arxiv.org/abs/2303.11074*

Huang, Y. 2021. "Technology innovation and sustainability: Challenges and research needs." *Clean Technologies and Environmental Policy*, *23* (6): 1663-1664.

ITU (International Telecommunication Union). 2025. *Facts and Figures 2025: Internet Use*. https://www.itu.int/itu-d/reports/statistics/2025/10/15/ff25-internet-use/

Jasanoff, S. 2016. *The Ethics of Invention: Technology and the Human Future*. WW Norton & Company.

Jensen, R. 2007. "The Digital Provide: Information (Technology), Market Performance, and Welfare in the South Indian Fisheries Sector." *The Quarterly Journal of Economics*, 122 (3): 879–923.

Jewell, C. 2019. "Artificial Intelligence: the New Electricity." *WIPO Magazine*, 13 June. Available at: https://www.wipo.int/en/web/wipo-magazine/articles/artificial-intelligence-the-new-electricity-55628 (Accessed: 2 July 2026).

Jonas, H. 1984. *The Imperative of Responsibility: In Search of an Ethics for the Technological Age*. University of Chicago Press.

Jones, B.F. and L.H. Summers, 2022. "A Calculation of the Social Returns to Innovation." In *Innovation and Public Policy*. University of Chicago Press: 13-60.

Kagan, S. 1989. *The Limits of Morality*. Oxford University Press.

Kahane, C. J. 2015. *Lives Saved by Vehicle Safety Technologies and Federal Motor Vehicle Safety Standards, 1960 to 2012.* (Report No. DOT HS 812 069).

Karnein, A., 2014. "Putting Fairness in its Place: Why There is a Duty to Take up the Slack." *The Journal of Philosophy*, *111* (11): 593-607.

Kasirzadeh, A. and I. Gabriel. 2025. "Characterizing AI Agents for Alignment and Governance." *arXiv preprint arXiv:2504.21848*.

Klosko, G. 2008. *Political Obligations*. Oxford University Press.

Korinek, A., and J. E. Stiglitz. 2021. "Artificial Intelligence, Globalization, and Strategies for Economic Development" *National Bureau of Economic Research Working Paper*, 28453 http://www.nber.org/papers/w28453

Kremer, M. 1998. "Patent Buyouts: A Mechanism for Encouraging Innovation." *The Quarterly Journal of Economics*, 113 (4): 1137-1167.

Krugman, P. 1979. “A Model of Innovation, Technology Transfer, and the World Distribution of Income.” *Journal of Political Economy*, 87 (2): 253-266.
Lippert-Rasmussen, K. 2016. *Luck Egalitarianism*. Bloomsbury Publishing.
Lanier, J. and E. G Weyl. 2018. “A Blueprint for a Better Digital Society.” *Harvard Business Review*, 26: 2-18.
Locke, J. 1689. *Two Treatises of Government*. N.p.
Lucas, R. E. “The Industrial Revolution: Past and Future.” *Lectures on Economic Growth* 109 (2002): 188.
Lyons, D. 1993. *Moral Aspects of Legal Theory: Essays on Law, Justice, and Political Responsibility*. Cambridge University Press.
Lythreatis, S., F. Acikgoz and N. Yassine. 2026. “Artificial Intelligence in Humanitarian Aid: A Review and Future Research Agenda.” *Technovation*, 151: 103415.
MacKenzie, D. and J. Wajcman, 1999. *The Social Shaping of Technology*. Open University.
Manyika, J. 2026. Introductory Notes: On AI, Science & the Future of Discovery. *Daedalus*, 155 (1-2): 7-33.
Maskus, K. E. 2000. *Intellectual Property Rights in the Global Economy*. Institute for International Economics.
Mazzucato, M. 2024. “Governing the Economics of the Common Good: From Correcting Market Failures to Shaping Collective Goals.” *Journal of Economic Policy Reform*, 27 (1): 1-24.
Merton, R. K. 1973. “The normative structure of science.” In N. W. Storer (Ed.), *The sociology of science: Theoretical and empirical investigations*. University of Chicago Press: (pp. 267–278). Original work published 1942.
Milanovic, B. 2011. *Worlds Apart: Measuring International and Global Inequality*. Princeton University Press.
Milanovic, B. 2015. “Global Inequality of Opportunity: How Much of Our Income is Determined by Where We Live?” *Review of Economics and Statistics*, 97 (2): 452-460.
Miller, D. 2004. “National Responsibility and Global Justice.” *Critical Review of International Social and Political Philosophy* 7 (1): 19–39.
Miller, D. 2007. *National Responsibility and Global Justice*. Oxford University Press.
Miller, R. D., 2004. “Beneficence, Duty and Distance.” *Philosophy & Public Affairs*, *32* (4):357-383.
Miller, S. and M.J. Selgelid, 2007. “Ethical and Philosophical Consideration of the Dual-Use Dilemma in the Biological Sciences.” *Science and Engineering Ethics,* 13 (4): 523-580.
Mills, C. W. 2005. ““Ideal theory” as Ideology.” *Hypatia*, 20 (3): 165-183.

Mohamed, S., M. T. Png, and W. Isaac, 2020. "Decolonial AI: Decolonial Theory as Sociotechnical Foresight in Artificial Intelligence." *Philosophy & Technology*, 33 (4): 659-684.

Mokyr, J. 1990. *The Lever of Riches: Technological Creativity and Economic Progress*. Oxford University Press.

Mokyr, J. 2025. *The Past and Future of Innovation: Can Progress be sustained?* (No. 2025-3). Nobel Prize Committee

Morozov, E. 2013. *To Save Everything, Click Here: The Folly of Technological Solutionism*. PublicAffairs.

Morsink, J. 1999. *The Universal Declaration of Human Rights: Origins, Drafting, and Intent*. University of Pennsylvania Press.

Muthukrishna, M., and J. Henrich. 2016. "Innovation in the collective brain." *Philosophical Transactions of the Royal Society B: Biological Sciences* 371, no. 1690.

Murphy, L. B. 2000. *Moral Demands in Nonideal Theory*. Oxford University Press.

Nagel, T. 2005. "The Problem of Global Justice." *Philosophy & Public Affairs* 33 (2): 113–47.

Narveson, J. 2003. "We Don't Owe Them a Thing! A Tough-Minded but Soft-Hearted View of Aid to the Faraway Needy." *The Monist*, *86* (3):419-434.

Nearing, G., D. Cohen, V. Dube, M. Gauch, O. Gilon, S. Harrigan, A. Hassidim, D. Klotz, F. Kratzert, A. Metzger, and A. Nevo. 2024. "Global Prediction of Extreme Floods in Ungauged Watersheds." *Nature*, *627*(8004), pp.559-563.

Nickel, J. W. 2007. *Making Sense of Human Rights.* 2nd ed. Blackwell.

Nordhaus, W. D. 1969. "An Economic Theory of Technological Change." *The American Economic Review*, 59 (2): 18-28.

Nordhaus, W. D., 2004. "Schumpeterian Profits in the American economy: Theory and Measurement."*National Bureau of Economic Research,* Working Paper 10433: 1-35

Nozick, R. 1974. *Anarchy, State, and Utopia*. Basic Books.

O'Keefe, C., P. Cihon, B. Garfinkel, C. Flynn, J. Leung and A. Dafoe. 2020. "The Windfall Clause." *Centre for the Governance of AI Research Report.* Future of Humanity Institute, University of Oxford.

O'Neill, O. 2000. *Bounds of Justice*. Cambridge University Press.

OPHI (Oxford Poverty and Human Development Index). 2025. "Global MPI Update 2025." OPHI. https://ophi.org.uk/global-mpi/2025.

Paine, T. 1797. *Agrarian Justice*. N.p.

Peter, O. and K. Devlin. 2025. "Decentralising LLM Alignment: A Case for Context, Pluralism, and Participation." In *Proceedings of the AAAI/ACM Conference on AI, Ethics, and Society*, 8 (2): 1988-1999.

Pettit, P. 1997. *Republicanism: A theory of freedom and government*. Oxford University Press.

Pogge, T. W. 2002. *World Poverty and Human Rights: Cosmopolitan Responsibilities and Reforms*. Polity Press.

Pomeranz, K. 2009. *The Great Divergence: China, Europe, and the Making of the Modern World Economy*. Princeton University Press.

Pope Leo XIV. (2026). *Magnifica humanitas* [Encyclical letter]. The Holy See. https://www.vatican.va/content/leo-xiv/en/encyclicals/documents/20260515-magnifica-humanitas.html

Poskett, J. 2022. *Horizons: A global history of science*. Penguin UK.

Prabhakaran, V., M. Mitchell, T. Gebru, T. and I, Gabriel. 2022. “A Human Rights-Based Approach to Responsible AI.” *arXiv preprint arXiv:2210.02667*.

Psacharopoulos, G. and H.A. Patrinos, 2018. “Returns to Investment in Education: a Decennial Review of the Global Literature.” *Education Economics*, *26* (5): 445-458.

Railton, P. 1984. “Alienation, Consequentialism, and the Demands of Morality.” *Philosophy & Public Affairs* 13 (2): 134–71.

Rawls, J. 1999. *A Theory of Justice*. Harvard University Press.

Rawls, J. 2001. *The Law of Peoples*. Harvard University Press.

Rawls, J. 2005. *Political Liberalism*. Columbia University Press.

Risse, M. 2012. *Global Justice: Arguments for the World*. Princeton University Press.

Robeyns, I. 2017. “Having Too Much.” *Nomos*, 58: 1-44.

Romer, P. M. 1994. “The Origins of Endogenous Growth.” *Journal of Economic perspectives*, 8 (1): 3-22.

Roser, M. 2019. “Global Inequality of Opportunity.” *Our World in Data*. https://ourworldindata.org/global-inequality-of-opportunity

Roser, M. 2021 . “The Economies That Are Home to the Poorest Billions of People Need to Grow If We Want Global Poverty to Decline Substantially.” *Our World in Data.* https://archive.ourworldindata.org/20260325-171315/poverty-growth-needed.html

Rotondi, V., R. Kashyap, L. M. Pesando, S. Spinelli and F. C. Billari. 2020. “Leveraging Mobile Phones to Attain Sustainable Development.” *Proceedings of the National Academy of Sciences*, *117* (24):13413-13420.

Ruggie, J. G. 2013. *Just Business: Human Rights in the Era of Market Power*. W. W. Norton & Company.

Samuelson, P.A, 1954. “The Pure Theory of Public Expenditure.” *The Review of Economics and Statistics*, *36* (4): 387-389.

Sauber-Schatz, E. K., D. J. Ederer, A. M. Dellinger, and G. T. Baldwin. 2016. “Vital Signs: Motor Vehicle Injury Prevention - United States and 19 Comparison Countries.” *MMWR. Morbidity and mortality weekly report* 65 (26): 672–677.

Scanlon, T.M., 1975. “Preference and Urgency.” *The Journal of Philosophy*, *72* (19): 655-669.

Scanlon, T. M. 1998. *What We Owe to Each Other*. The Belknap Press of Harvard University Press.

Scheffler, S. 1994. *The Rejection of Consequentialism: A Philosophical Investigation of the Considerations Underlying Rival Moral Conceptions*. Oxford University Press.

Scheffler, S. 2002. *Boundaries and Allegiances: Problems of Justice and Responsibility in Liberal Thought*. OUP Oxford.

Sen, A. 1983. “Poor, relatively speaking.” *Oxford Economic Papers*, 35 (2):153-169.

Sen, A. 2004. “Elements of a Theory of Human Rights.” *Philosophy & Public Affairs* 32 (4): 315–56.

Sen, A. 2009. *The Idea of Justice*. Belknap Press.

Shaver, L. 2015. “The Right to Science: Ensuring that Everyone Benefits from Scientific and Technological Progress.” *European Journal of Human Rights*, 4: 411-430.

Shevlane, T., S. Farquhar, B. Garfinkel, M. Phuong, J. Whittlestone, J. Leung, D. Kokotajlo, N. Marchal, M. Anderljung, N. Kolt, L. Ho et al. 2023. “Model Evaluation for Extreme Risks.” *arXiv preprint arXiv:2305.15324*.

Shue, H. 1980. *Basic Rights: Subsistence, Affluence, and U.S. Foreign Policy*. Princeton University Press.

Shue, H. 1988. “Mediating Duties.” *Ethics*, 98 (4): 687–704.

Simmons, A.J. 1999. “Justification and legitimacy.” *Ethics*, *109* (4): 739-771.

Simmons, A.J. 2020. *Moral Principles and Political Obligations*. Princeton University Press.

Singer, P. 1972. “Famine, Affluence, and Morality.” *Philosophy & Public Affairs* 1 (3): 229–43.

Smart, J. J. C., and B. Williams. 1973. *Utilitarianism: For and Against*. Cambridge University Press.

Smith, A. 1776. *The wealth of nations*. N.p.

Solow, R. M. 1957. “Technical Change and the Aggregate Production Function.” *The Review of Economics and Statistics* 39 (3): 312–20.

Steiner, H. 1994. *An Essay on Rights*. Wiley Publishing.

Stemplowska, Z. 2016. “Doing More than One’s Fair Share.” *Critical Review of International Social and Political Philosophy*, 19 (5): 591-608.

Stern, N., M. Romani, R. Pierfederici, M. Braun, D. Barraclough, S. Lingeswaran, E. Weirich-Benet, and N. Niemann, 2025. “Green and Intelligent: the Role of AI in the Climate Transition.” *npj Climate Action* 4, 56: 1-7.

Strauss, J. and Thomas, D. 1998. "Health, Nutrition, and Economic Development." *Journal of Economic Literature*, 36 (2): 766-817.

Tan, K. C. 2004. *Justice Without Borders: Cosmopolitanism, Nationalism, and Patriotism*. Cambridge University Press.

Tasioulas, J. 2015. "On the Foundations of Human Rights." In *Philosophical Foundations of Human Rights*, edited by R. Cruft, M. Renzo, and S. Liao. Oxford University Press.

Temkin, L. S. 1993. *Inequality*. Oxford University Press.

Tørsløv, T., L. Wier, L. and G. Zucman. 2023. "The Missing Profits of Nations." *The Review of Economic Studies*, 90 (3): 1499-1534.

UBS. 2025. *Global Wealth Report 2025*. UBS Global Wealth Management. https://www.ubs.com/global/en/wealthmanagement/insights/global-wealth-report.html.

UNICEF (United Nations Children's Fund). (2026). "Under-Five Mortality." Updated March. https://data.unicef.org/topic/child-survival/under-five-mortality/.

United Nations. 2024. *Governing AI for Humanity.* United Nations. https://www.un.org/sites/un2.un.org/files/governing_ai_for_humanity_final_report_en.pdf

United Nations Development Programme (UNDP). 2025. *Human Development Report 2025: A Matter of Choice: People and Possibilities in the Age of AI*. UNDP.

United Nations General Assembly, 1948. *Universal Declaration of Human Rights.* United Nations General Assembly.

United Nations General Assembly, 1966. *International Covenant on Economic, Social and Cultural Rights*. United Nations General Assembly.

Vallentyne, P. 2000. *Left-libertarianism: A Primer*. Palgrave-Macmillan.

Van Dijk, J. A. 2005. *The Deepening Divide: Inequality in the Information Society*. Sage publications.

Waldron, J. 1989. "Rights in Conflict." *Ethics*, 99 (3): 503-519

Waldron, J. 1990. *The Right to Private Property*. Oxford University Press.

Warschauer, M. 2004. *Technology and Social Inclusion: Rethinking the Digital Divide*. MIT Press.

Wenar, L, 2008. "Property Rights and the Resource Curse." *Philosophy & Public Affairs*, 36 (1): 2-32.

Wenar, L., 2015. *Blood Oil: Tyrants, Violence, and the Rules that Run the World*. Oxford University Press.

Winner, L. 2020. *The Whale and the Reactor: A Search for Limits in an Age of High Technology*. University of Chicago Press.

Woods, N. 2006. *The Globalizers: The IMF, the World Bank, and Their Borrowers*. Cornell University Press.

World Bank, 2016. *World Development Report 2016: Digital Dividends*. World Bank.
World Bank, 2025. “September 2025 Global Poverty Update from the World Bank: New Data and Regional Classifications.” *World Bank Blog*, September 30. . https://blogs.worldbank.org/en/opendata/september-2025-global-poverty-update-from-the-world-bank--new-da.
Young, I. M. 2011. *Responsibility for Justice*. Oxford University Press.
Zufiria, P. J., D. Pastor-Escuredo, L. Úbeda-Medina, M. A. Hernandez-Medina, I. Barriales-Valbuena, A. J. Morales, D. C. Jacques, W. Nkwambi, M. B. Diop, J. Quinn, and P. Hidalgo-Sanchís. 2018. “Identifying Seasonal Mobility Profiles from Anonymized and Aggregated Mobile Phone Data. Application in Food Security.” *PLoS ONE* 13 (4): e0195714.